\documentclass[12pt]{iopart}

\usepackage{graphicx}
\usepackage{dcolumn}
\usepackage{bm}
\usepackage{color}

\begin{document}

\title{Magnonic crystals for data processing}
\author{A V Chumak, A A Serga, and B Hillebrands}

\address{Fachbereich Physik and Landesforschungszentrum OPTIMAS, Technische Universit\"at Kaiserslautern, 67663
Kaiserslautern, Germany}
\ead{chumak@physik.uni-kl.de}

\begin{abstract}

Magnons---the quanta of spin waves---propagating in magnetic materials with wavelengths at the nanometer-scale and carrying information in the form of an angular momentum, can be used as data carriers in next-generation, nano-sized low-loss information processing systems. In this respect, artificial magnetic materials with properties periodically varied in space, known as magnonic crystals, are especially promising for controlling and manipulating the magnon currents. In this article, different approaches for the realization of static, reconfigurable, and dynamic magnonic crystals are presented along with a variety of novel wave phenomena discovered in these crystals. Special attention is devoted to the utilization of magnonic crystals for processing of analog and digital information. 

\end{abstract}


\maketitle

\section{Introduction}

This article focuses on a selection of results in the field of magnonic crystals obtained within the last decade in our group. Special attention is devoted to the application of magnonic crystals for data processing and information technologies. 
Because of the large number of achievements in the field, it is unavoidable that many important results are out of scope of this article.  
Therefore, we would like to attract the reader's attention to the excellent reviews devoted to different aspects of magnonic crystals: Spin-wave dynamics in periodic structures for microwave applications \cite{MC-review1, MC-review2}, Brillouin light scattering studies of planar metallic magnonic crystals \cite{24-MC-review}, micromagnetic computer simulations of width-modulated waveguides \cite{25-kim}, photo-magnonic aspects of the antidot lattices studies \cite{15-blocksMagnonics}, theoretical studies of one-dimensional monomode waveguides \cite{26-MC}, and reconfigurable magnonic crystals \cite{27-MC-review}. The results presented in the current review were partially presented already in our own reviews ``\emph{YIG magnonics}'' \cite{A1-serga-2010-JPD-YIG} and ``\emph{Magnon spintronics}'' \cite{A3-chumak-2015-NP-MagSpin} as well as in book chapters on dynamic magnonic crystals \cite{A15-chumak-book-2012-DMC} and magnon spintronics \cite{book-magnonspintr}. 

The article is organised in the following way: In the introductory section we present the field of magnon spintronics and discuss the role of magnonic crystals in it. The next Section 2 is devoted to the properties of spin waves in thin planar films and waveguides, to the magnetic materials commonly used in the field, and to the methods of spin-wave excitation and detection.  Section 3 and 4 are devoted to the study of physics phenomena taking place in static and dynamic magnonic crystals, respectively. The application of magnonic crystals for magnon-based processing of analog and digital data is discussed in Section 5. Specifically, the static magnonic crystals can be used as passive elements like microwave filters, resonators or delay lines for microwave generation. Reconfigurable and dynamic magnonic crystals are promising as an active device performing, for example, tunable filtering, frequency conversion or time reversing. In the final section we briefly summarize the results. 

\subsection{Magnon spintronics: From electron- to magnon-based computing}

A disturbance in the local magnetic order can propagate in a magnetic material in the form of a wave. This wave was first predicted by F. Bloch in 1929 and was named spin wave since it is related to a collective excitation of the electron spin system in ferromagnetic metals and insulators \cite{1-bloch}. Since that time, the field of magnetization dynamics has grown into a wide domain of science in which the studies of coherent externally-driven spin waves are of particular importance. The spin-wave characteristics can be engineered by a wide range of parameters including the choice of the magnetic material, the shape of the sample as well as the orientation and size of the applied biasing magnetic field \cite{2-gurevich, 3-stancil2}. This, in combination with a rich choice of linear and non-linear spin-wave properties, makes spin waves to be an excellent object for studies of general wave physics \cite{A1-serga-2010-JPD-YIG}. One- and two-dimensional soliton formation \cite{4-Soliton-1, 5-bullet-2}, non-diffractive spin-wave caustic beams \cite{6-caustics, 7-caustics-2, 8-caustics-3}, wave-front reversals \cite{9-WFR-melkov2, 10-WFR_2D}, room temperature Bose-Einstein condensation of magnons \cite{11-BEC, 12-BEC-Serga}, and formation of magnon supercurrents \cite{supercurrent} is just a small selection of examples. 

On the other hand, spin waves in the GHz frequency range are of high interest for applications in telecommunication systems and radars since the spin-wave wavelengths are orders of magnitude smaller than those of electromagnetic waves opening access to the miniaturisation of analog data processing elements. Nowadays, spin waves and their quanta, magnons, attract much attention also due to another very ambitious perspective: They are being considered as data carriers in novel computing devices instead of electrons in electronics. The main advantages offered by magnons for data processing are \cite{A3-chumak-2015-NP-MagSpin, A2-stamps-2014-JPD-Roadmap}:

\begin{itemize}
  \item{Potential to realize novel, highly efficient, wave-based computing concepts.}
\item{Spin waves can be used in devices of sizes down to sub-10 nm.}
\item{Magnon frequencies cover a very wide range from sub-GHz to tens of THz.}
\item{Magnons allow for room-temperature transport of spin information without the translational motion of electrons and, therefore, without the generation of parasitic Joule heat.}
\item{Pronounced spin-wave nonlinear effects allow for the realization of magnon-magnon interaction-based functionalities, such as signal gating in a magnon transistor.}
\end{itemize}

\begin{figure}
\begin{center}
\includegraphics[width=0.5\columnwidth]{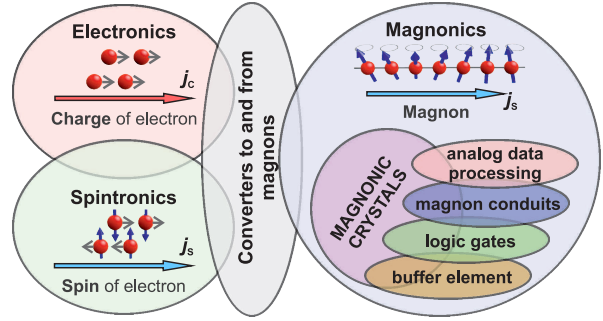}
\caption{\label{Fig1} The concept of magnon spintronics. Data coded in charge- or spin-currents is converted to magnon currents, processed within the magnonic system and converted back. The magnonic crystal is a universal element in the field of magnonics which can be used for data transfer, processing, and buffering. }
\end{center}
\end{figure}

The field of science that refers to information transport and processing by spin waves is known as magnonics \cite{15-blocksMagnonics, A1-serga-2010-JPD-YIG, 13-Kruglyak_01, 14-magnonics}. The utilization of magnonic approaches in the field of spintronics, hitherto addressing electron-based spin currents, gave birth to the field of magnon spintronics \cite{A3-chumak-2015-NP-MagSpin, A2-stamps-2014-JPD-Roadmap}. The diagram of magnon spintronics in Fig.~1 shows that, besides magnon-based elements operating with analog and digital data, this field also comprises converters between the magnon subsystem and electron-based spin and charge currents. Different approaches including utilisation of spin transfer torque, spin Hall effect, magneto-acoustic phenomena, opto-magnetic phenomena, and many others are under intensive investigations nowadays. These convertors are outside the scope of the article but are partially presented in the review \cite{A3-chumak-2015-NP-MagSpin}.
 
\subsection{Magnonic crystals and their role in the field of magnon spintronics}
 
Magnonic crystals, which are artificial magnetic media with a periodic lateral variation of their properties, are of high interest for both pure wave physics and application oriented magnonics. Spectra of spin-wave excitations in such structures are significantly different compared to uniform media and exhibit features such as band gaps, where spin waves are not allowed to propagate. In spite of the fact that the term ``magnonic crystal'' is relatively new (it was first introduced by Nikitov \textit{et al.} in 2001 \cite{16-MC}), this field goes back to studies of spin-wave propagation in periodical structures which have been initiated by Sykes, Adam and Collins already in 1976 \cite{17-MC-grooves1}. The early magnonic crystal research activities were mainly restricted by their application to the development of microwave filters and resonators -- see reviews \cite{MC-review1, MC-review2, 20-IEEE-4}. 
 
Nowadays, when linear and non-linear coupling of spin-wave modes and demagnetizing effects cannot be neglected on the nano- and micro-scales, many studies are focused on the understanding of spin-wave related physical phenomena in magnonic crystals. In particular, high attention attracted investigations of magnonic crystals with defects \cite{defect-1, defect-2, defect-3, defect-4, defect-5}, topological and nonreciprocal phenomena in magnonic crystals \cite{nonrecipr-1, nonrecipr-2, nonrecipr-3, nonrecipr-4, nonrecipr-5, nonrecipr-6, nonrecipr-7, nonrecipr-8, nonrecipr-9}, linear and non-linear spin-wave dynamics in coupled magnonic crystals \cite{nikitov-coupler, coupled-1, coupled-2}, and formation and propagation of solitons \cite{21-MCSoliton, loop-1, soliton-1, loop-3}. Moreover, besides one-dimensional magnonic crystals \cite{1D MC-1, 1D MC-2, 1D MC-3, 1D MC-4, 1D MC-5, 1D MC-6, 1D MC-7, 1D MC-8, 1D MC-9, 1D MC-10, 1D MC-11, 1D MC-12, 1D MC-13, 1D MC-14, 1D MC-15, 1D MC-16, 1D MC-17, 1D MC-18}, two-dimensional magnonic crystals \cite{2D MC-1, 2D MC-2, 2D MC-3, 2D MC-4, 2D MC-5, 2D MC-6, 2D MC-7, 2D MC-8, 2D MC-9, 2D MC-10, 2D MC-11, 2D MC-12, 2D MC-13, 2D MC-14, 2D MC-15, 2D MC-16, 2D MC-17, 2D MC-18, 2D MC-19, 2D MC-20, 2D MC-21, 2D MC-22, 2D MC-23, 2D MC-24, 2D MC-25, 2D MC-26} have been intensively studied experimentally while three-dimensional magnonic crystals \cite{23-MC-theory2, 3D MC-1, 3D MC-2} are investigated theoretically. Thus, the magnonic crystal field is growing fast. For magnon spintronic applications, magnonic crystals constitute one of the key elements since they open access to novel multi-functional magnonic devices \cite{A3-chumak-2015-NP-MagSpin}. These devices can be used as spin-wave conduits and filters (we do not provide any citations here since, in fact, any magnonic crystal can serve as a conduit or a filter), sensors \cite{sensor-1, 105-SAW-1, sensor-2}, delay lines and phase shifters \cite{A18-karenowska-2010-APL-Gener, loop-2}, components of auto-oscillators \cite{loop-1, A18-karenowska-2010-APL-Gener, loop-3, loop-2}, frequency- and time-inverters \cite{A19-chumak-2010-NC-TimeRev, A20-karenowska-2012-PRL-DMC}, data buffering elements \cite{A20-karenowska-2012-PRL-DMC, A21-chumak-2012-PRL-storage}, power limiters \cite{22-MCNonlinear}, non-linear enhancers in a magnon transistor \cite{A23-chumak-2014-NC-transist}, and components of logic gates \cite{122-khitun3, A16-nikitin-APL-2015-DMC}. 

\section{Essentials of spin-wave dynamics}

\subsection{Spin waves in thin magnetic films and waveguides}
 
Spin waves are related to a collective excitation of the electron spin system in ferromagnetic metals and insulators (see books \cite{2-gurevich, 3-stancil2}). The main spin-wave characteristics can be obtained from the analysis of its dispersion relation, i.e. the dependence of the wave frequency $f$ on its wavenumber $k$. There are two main contributors to the magnon energy: long-range dipole-dipole and short-range exchange interactions. As a result, the dispersions of spin waves are significantly different from the well-known dispersion of light or sound in uniform media. Moreover, spin-wave dispersion relations in in-plane magnetized films are strongly anisotropic due to the given orientation of the magnetization of the magnetic medium. Therefore, for example, spin waves propagating along the magnetization direction have a different dispersion in comparison with the waves propagating transverse to the magnetization (see Fig.~2) \cite{31-damon-eshbach2}.
 
In most practical situations spin waves are studied in spatially confined samples such as thin films or strips magnetized in-plane by an external magnetic field. The geometry of a spin-wave waveguide, namely its thickness and its width, is a key parameter defining the spin-wave dispersion along with the magnetic properties of the material and the applied magnetic field. In order to underline the role of the waveguide geometry, the dispersion relations for infinite plane films of different thickness as well as for a spin-wave waveguide are shown in Fig.~2.
 
\begin{figure}
\begin{center}
\includegraphics[width=0.9\columnwidth]{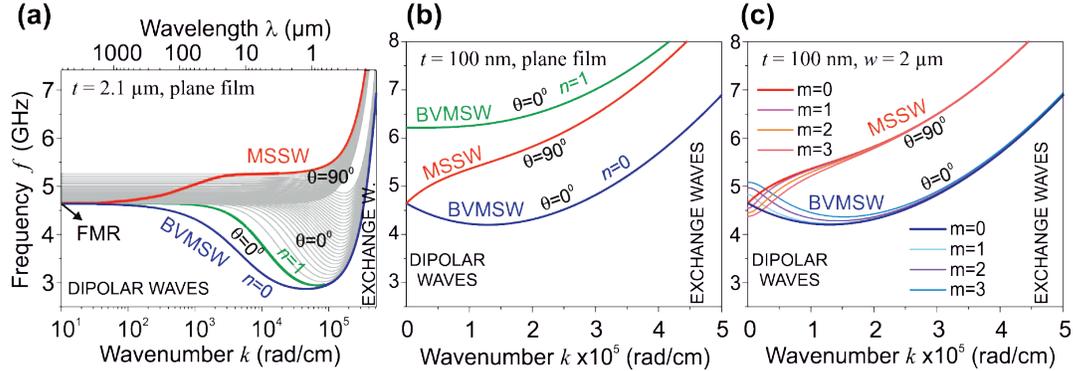}
\caption{\label{Fig2} Spin-wave dispersion characteristics for an infinite in-plane magnetized Yttrium Iron Garnet (YIG) film of 2.1 $\mu$m thickness (a), infinite film of 100 nm thickness (b), and 2 $\mu$m wide spin-wave waveguide of 100 nm thickness (c). A magnetic field of 100 mT is applied in-plane, saturation magnetization 140 kA/m, exchange constant 3.6 pJ/m, $\theta$ is the angle between the spin-wave propagation direction and magnetization. (a) Logarithmic scale is used for the wavenumber. The blue line shows the lowest $n = 0$ thickness mode of a backward volume magnetostatic wave (BVMSW) propagating along the magnetic field; the green line shows the first $n = 1$ BVMSW thickness mode and the red line shows the magnetostatic surface spin wave (MSSW) propagating transversely to the orientation of the magnetic field. 29 higher order thickness modes are shown in grey. The dispersions were found by numerical calculations  using the approach developed in Refs.\,\cite{32-kalinikos-iee, 33-kalinikos-slavin, 34-kalinikos-slavin-2}. (b) A linear scale is used for the wavenumber. Dispersions for MSSW as well as zero $n = 0$ and first $n = 1$ BVMSW modes are shown. (c) Dispersion for zero thickness mode $n = 0$ but different width modes $m$ are shown. Calculations in (b) and (c) are performed using simplified analytical expressions in \cite{33-kalinikos-slavin}, with additional account of width modes $k_\mathrm{SUM}^2 = k^2 + (m\pi/w)^2$ and the dependence of the spin-wave propagation angle on the width mode $\phi(w, m)$.}
\end{center}
\end{figure}

The typical spin-wave dispersion characteristics for an in-plane magnetized YIG film of micrometer thickness are shown in Fig.~2a. In spite of the fact that spin waves can propagate at any angle $\theta$ relative to the magnetization orientation, only the dispersions for the cases of the longitudinal ($\theta = 0^{\circ}$) and transverse ($\theta = 90^{\circ}$) wave propagation are shown for simplicity. The spectrum comprises three main regions: the region of small wavenumbers $k < 10^4$ rad/cm corresponds to dipolar waves usually termed magnetostatic waves (MSW), the region of large $k > 10^5$ rad/cm corresponds to exchange waves, and the region in between corresponds to dipolar-exchange spin waves (DESWs). One can see that, as opposed to the case presented in Fig.~2b and Fig.~2c, the dispersions for the micrometer thick YIG film show: (i) a large decrease in the frequency of the backward volume magnetostatic wave (BVMSW) mode to well below the ferromagnetic resonance (FMR) frequency, and (ii) a large density of higher-order thickness BVMSW modes which are shown in grey. Please note that the magnetostatic surface spin wave (MSSW) mode has a hyperbolic rather than a cosine distribution of dynamic magnetization across the film thickness and, therefore, does not possess higher-order thickness modes. The complexity of the spectrum plays a crucial role in strongly overpopulated non-linear magnonic systems and, for example, might result in the formation of BEC of magnons \cite{11-BEC, 12-BEC-Serga}, generation of magnon supercurrents \cite{supercurrent}, and in effective nonlinear magnon scattering in a magnon transistor \cite{A23-chumak-2014-NC-transist}.

In contrast, the spin-wave spectrum of a thin film is significantly diluted, and only the first thickness mode is shown in Fig.~2b (the other thickness modes not shown here have much higher frequencies). This is due to the fact that quantization across the film thickness results in an increase in the spin-wave frequency caused by exchange interaction. Figure\,2c shows how the spin-wave quantization along the width of a waveguide changes the dispersion relations. One can see that the quantization of the modes across the waveguide results in a pronounced modification of the dispersion characteristics in the dipolar region. The short-wavelength exchange waves are much less sensitive to the geometry of the waveguide.
 
\subsection{Magnetic materials for magnonic applications}
 
 As was discussed in the previous Section, spin waves are usually studied in thin magnetic films or waveguides fabricated in the form of narrow magnetic strips. The choice of the magnetic material plays a crucial role in fundamental as well as in applied magnonics. The main requirements are: (i) small Gilbert damping parameter $\alpha$ in order to ensure long spin-wave lifetimes; (ii) large saturation magnetization for high spin-wave frequencies and velocities; (iii) high Curie temperatures to provide thermo-stability; and (iv) simplicity in the fabrication of magnetic films and in the patterning processes. 
 
The most commonly used materials for magnonics as well as those with a high potential for magnonic applications are presented in Table 1 together with some selected parameters and estimated spin-wave characteristics. These characteristics are: spin-wave lifetime $\tau$ defined, first of all, by the Gilbert damping parameter $\alpha$; spin-wave group velocity $v$ that, together with the lifetime define the spin wave $1/e$ propagation distance (mean free path) $l = \tau v$; and the ratio of the mean free path to the wavelength $l/\lambda$. The latter is especially important for magnonic applications and its increase is one of the primary challenges facing the field \cite{A3-chumak-2015-NP-MagSpin}. Since practically all experimental studies in magnonics are performed nowadays for dipolar waves, the characteristics in the table are estimated using the dipolar approximation where the exchange interaction is ignored. However, it has to be mentioned that the use of short-wavelength exchange waves appears to be promising in particularly from the point of view of an increase in the $l/\lambda$ ratio.

\begin{table}
\begin{center}
\includegraphics[width=0.9\columnwidth]{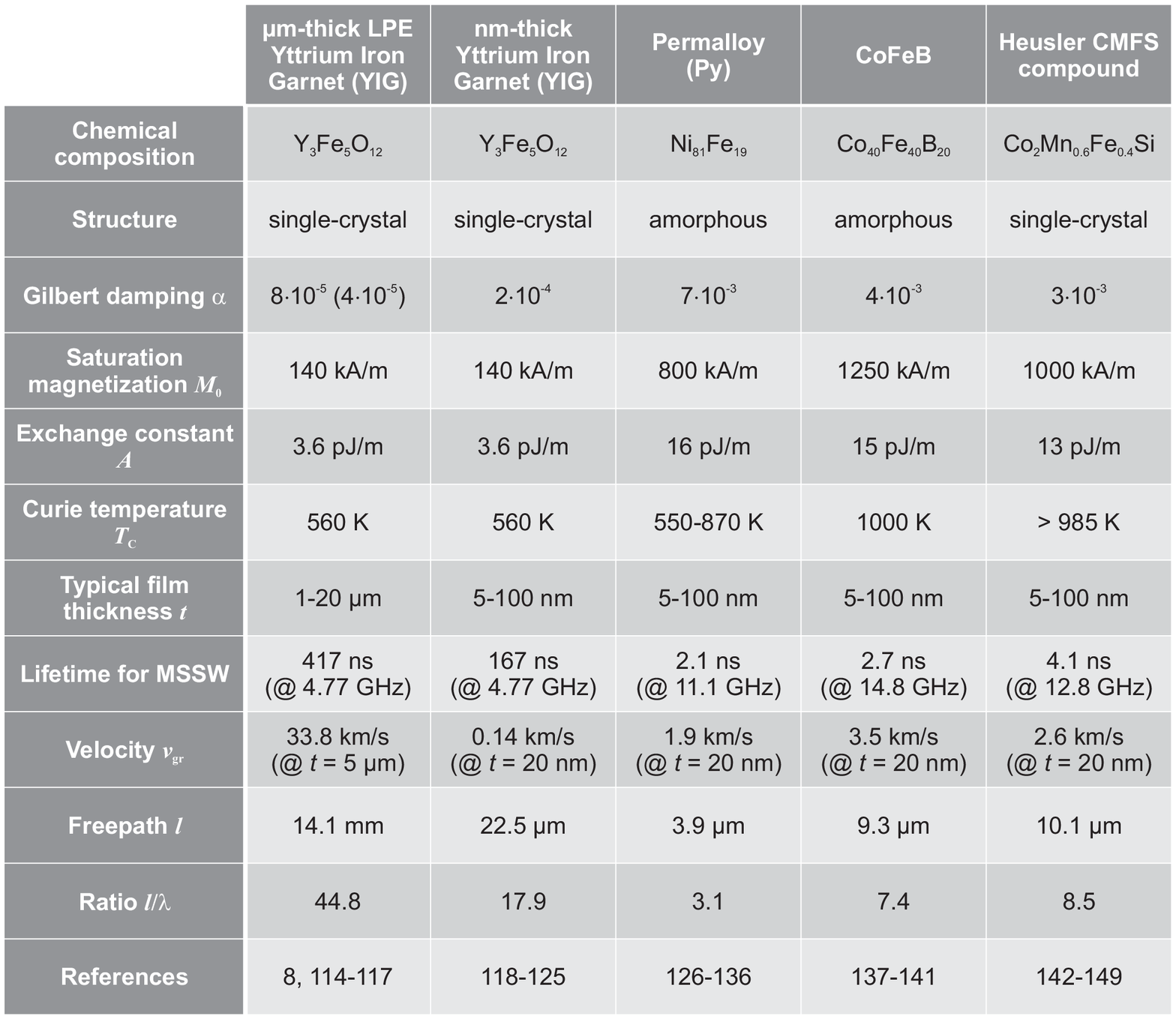}
\caption{\label{Table1} Selection of magnetic materials for magnonic applications, their main parameters, and estimated spin-wave characteristics. The characteristics are calculated using the dipolar approximation for infinite films magnetized in-plane with a 100 mT magnetic field. Magnetostatic surface spin waves that propagate perpendicularly to the magnetization direction and have the largest group velocities are considered. For simplicity, the spin wave of wavenumber $k = 0.1/t$ is analysed. Lifetime is estimated using the approach of bulk material without taking ellipticity into account \cite{3-stancil2}.}
\end{center}
\end{table}
 
The first material in the table is a monocrystalline Yttrium Iron Garnet Y$_3$Fe$_5$O$_{12}$ (YIG) film grown by high-temperature liquid phase epitaxy (LPE) on Gadolinium Gallium Garnet (GGG) substrates \cite{35-YIG-1, 36-YIG-2, 37-IEEE-2, 38-YIG-saga}. This ferrimagnet has the smallest known magnetic loss that results in a spin-wave lifetime of some hundreds of nanoseconds and, therefore, finds widespread use in academic research \cite{A1-serga-2010-JPD-YIG}. Most of the experimental results presented in this thesis were obtained using LPE YIG. The small magnetic loss is due to the fact that YIG is a magnetic dielectric (ferrite) with very small spin-orbit interaction and, consequently, with small magnon-phonon coupling \cite{38-YIG-saga}. Moreover, high quality of LPE single-crystal YIG films ensure a small number of inhomogeneities and, thus, suppressed two-magnon scatterings \cite{2-gurevich, 66-sparks}. However, the thickness of these films, which is in the micrometer range, does not allow for the fabrication of YIG structures of nanometer sizes. Therefore, the fabrication of nanostructures became possible only within the last few years with the development of technologies for the growth of high-quality nm-thick YIG films, see second column in Table 1, by means of e.g. pulsed-laser deposition (PLD) \cite{39-nanoYIG, 40-nanoYIG-2, 41-nanoYIG-3, 42-nanoYIG-4, A5-onbasli-2014-APLM-YIG}, sputtering \cite{43-nanoYIG-5}, or via modification of the LPE growth technology \cite{A4-pirro-2014-APL-microYIG, LPE-YIG-Dubs}. Although the quality of these films is still worse when compared to micrometer-thick LPE YIG films, it is already good enough to satisfy many requirements of magnonic applications \cite{A3-chumak-2015-NP-MagSpin}. 
 
The second most commonly used material in magnonics is Permalloy which is a polycrystalline alloy of 80$\%$ Ni - 20$\%$ Fe (see Table 1). This is a soft magnetic material with low coercivity and anisotropies. One of the major advantages of this material is that it has a fairly low spin-wave damping value considering it is a metal, and it can be easily deposited and nano-structured \cite{44-Py-1, 45-patton-permalloy-damping, 46-FMR, 47-ANTENNA, 48-nanoYIG-3, 49-multiplexer, 50-micropumping, A9-chumak-2009-APL-muMC}. Therefore, Permalloy was intensively used for the investigation of spin-wave physics in micro-structures (see reviews \cite{51-BLS, 52-mesoscale}) and for the investigations of magnonic crystals in particular \cite{24-MC-review, 15-blocksMagnonics, 27-MC-review, A9-chumak-2009-APL-muMC, A12-obry-2013-APL-ionMC}. Nowadays, much attention of the community is also focused on CoFeB \cite{53-CoFeB-1, 54-CoFeB-2, 55-CoFeB-3, 56-CoFeB-4, 57-CoFeB-5} and half-metallic Heusler compounds \cite{58-Heusler-1, 59-Heusler-2, 60-Heusler-3, 61-Heusler-4, 62-Heusler-5, 63-Heusler-6, 64-Heusler-7, 65-Heusler-8}. These materials possess smaller Gilbert damping parameters and larger values of saturation magnetization, and, therefore, are more suitable for the purposes of magnonics. For example, it was demonstrated that the spin-wave mean free path in Heusler compounds can reach 16.7 $\mu$m \cite{64-Heusler-7}. 
 
The fabrication of high-quality spin-wave waveguides is also one of the primary tasks in the field of magnonics. The most commonly used technique for the fabrication of micrometer-thick YIG waveguides in the form of magnetic strips is a dicing saw \cite{67-MSSW-1} since the width of the waveguide is usually larger than 1~mm. The main technique for the patterning of such YIG films is photolithography with subsequent wet etching by means of hot orthophosphoric acid \cite{A6-chumak-2008-APL-MC}. At the same time the recently proposed laser-based patterning of YIG films shows large potential \cite{daimon-2015}. Another technique has been used to pattern nanometer-thick YIG films: e-beam lithography with subsequent Ar$^{+}$ dry etching have shown good results \cite{42-nanoYIG-4, A4-pirro-2014-APL-microYIG}. Focused ion beam (FIB) milling has also recently shown very promising results - a 70 nm wide spin-wave waveguide was fabricated into a 100\,nm thick YIG film at the Nano-Structuring Center at the University of Kaiserslautern (not published). The same techniques can also be used for the patterning of metallic magnetic films. However, more often a different approach is used: The magnetic material is deposited on a resist mask produced via photo- or electron beam lithography followed by a standard lift-off process. Antennas and the required contact pads are deposited afterwards in subsequent lithography and lift off processes.

\subsection{Methods of spin-wave excitation and detection}

Modern magnonics consists of a wide range of instrumentation for the excitation and detection of magnons. The most commonly used techniques as well as techniques showing large potential are given in Table 2 with references and brief descriptions of their particular features. It can be seen that three main categories are: microwave, optical, and spintronics approaches. The main requirements for magnon detection techniques could be defined as sensitivity, the range of detectable wavelengths and frequencies, as well as frequency, spatial, and temporal resolution. For the spin-wave excitation techniques the efficiency of excitation, its coherency as well as the wavenumber range are of primary importance.
\cite{68-MSSW-2, 69-schneider, 70-wireless, 71-anti-dot, 72-YIG-FMR, 73-kennenwell, 74-smith, 75-BLS-YIG, 76-ultrafast, 77-optic, 78-bauer, 79-spin pumping, costache, 80-kajiwara, 81-SHE, 82-STT-1, 83-STT-2, 84-STT-3, 85-STT-4, 86-STT-5, 87-STT-6, 88-STT-7, van Wees, STT-8, 89-plihal, 90-tang, 91-bus, 92-dutta, 93-FMRFM, 94-an, 95-NRS, 96-xray, 97-xray, 98-xray, 99-electron, 100-electron}

The conventional technique for spin-wave excitation is inductive spin-wave excitation with a microwave current sent through a strip-line antenna. In order to understand the spin-wave signal excitation mechanism, it is useful to consider the waveguide as a reservoir of quasi-classical spins. When the waveguide is magnetically saturated, the mean precessional axis of all the spins is parallel to the bias field. Application of a microwave signal to the strip-line antenna generates an alternating Oersted magnetic field around it. The components of this field which are perpendicular to the bias direction create a torque on the magnetization which results in an increase in the precessional amplitude \cite{A1-serga-2010-JPD-YIG, 68-MSSW-2, 69-schneider}. Spins that precess under the antenna interact with their nearest neighbors. If the correct conditions for field and frequency are satisfied, spin-wave propagation is supported. After propagation, the spin-wave excitation might be detected by a similar output antenna. The mechanism of spin-wave detection is, by symmetry, the inverse of the excitation process. 
 
\begin{table}
\begin{center}
\includegraphics[width=0.9\columnwidth]{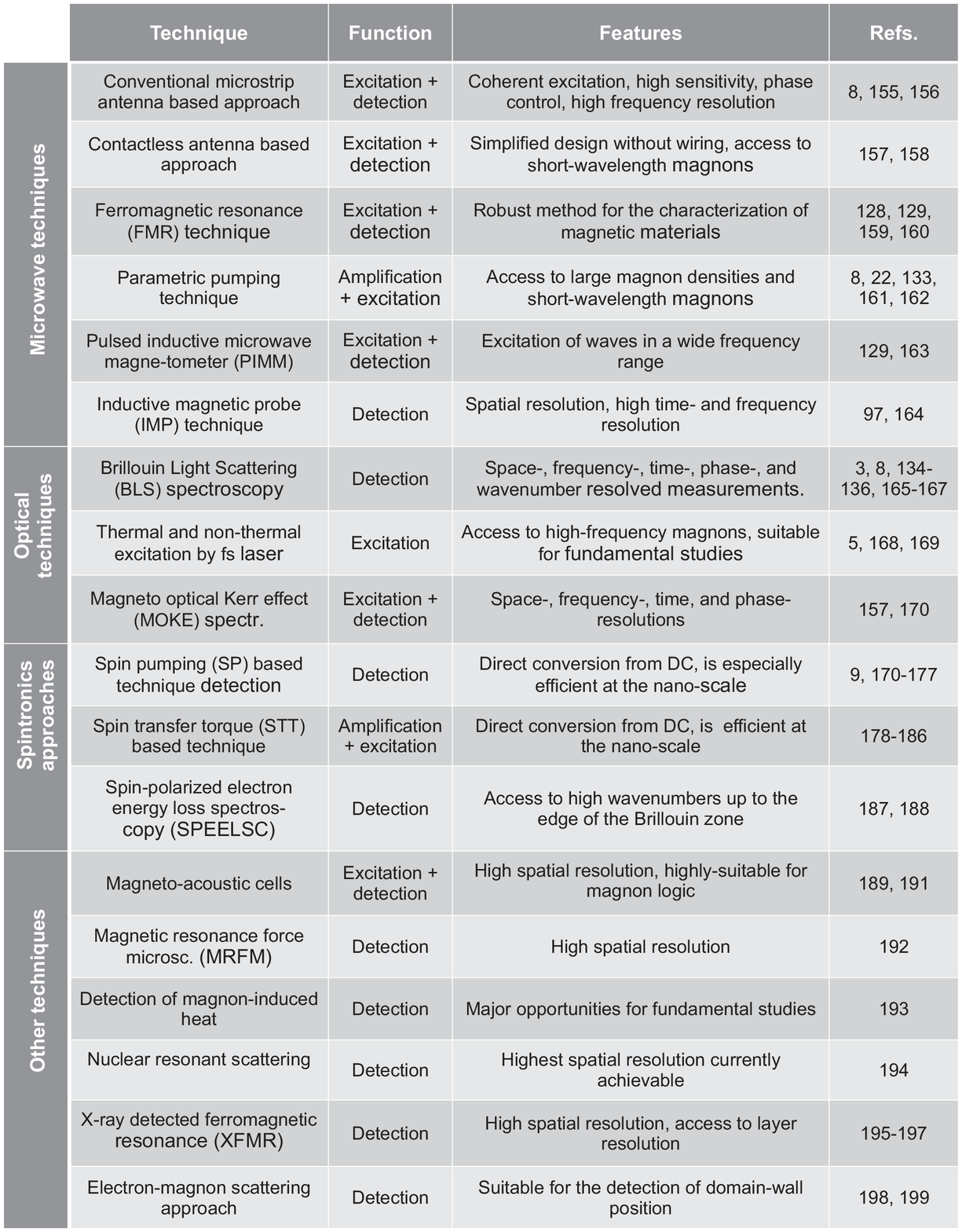}
\caption{\label{Table2} Techniques for excitation and detection of spin waves and main features.}
\end{center}
\end{table}
 
Number one of other detection techniques in magnonics nowadays is probably Brillouin Light Scattering (BLS) spectroscopy \cite{A1-serga-2010-JPD-YIG, 51-BLS, 52-mesoscale, 75-BLS-YIG}. The physical basis of BLS spectroscopy is the inelastic scattering of photons by magnons. Scattered light from a probe beam, incident on the sample is analyzed and allows the frequencies and wavenumbers of the scattering magnons to be determined, where the scattered photon intensity is proportional to the spin-wave intensity. The technique is generally used in conjunction with a microwave excitation scheme and, over the last decade, has undergone extensive improvements. BLS spectroscopy now achieves a spatial resolution of 250 nm, and time-, phase-, and wavenumber resolved BLS spectroscopy has been realized \cite{24-MC-review, A1-serga-2010-JPD-YIG, A9-chumak-2009-APL-muMC, 51-BLS, 52-mesoscale, 75-BLS-YIG, 101-phaseBLS, 102-kBLS}.
 
\section{Static magnonic crystals}
\subsection{Spin-wave Bragg scattering conditions in magnonic crystals}
 
Magnonic crystals, whose name refers to the spin-wave quasiparticles---magnons, are artificial magnetic media with periodic variation of their magnetic properties in space \cite{16-MC}. Similarly to photonic crystals operating with light, magnonic crystals use the wave nature of magnons to obtain magnon propagation characteristics that are inaccessible by any other means (see reviews \cite{24-MC-review, 25-kim, 15-blocksMagnonics, 26-MC, 27-MC-review, A1-serga-2010-JPD-YIG, A3-chumak-2015-NP-MagSpin}). Bragg scattering affects a spin-wave spectrum in such a periodic structure and leads to the formation of band gaps - frequencies at which spin-wave propagation is prohibited.

\begin{figure}
\begin{center}
\includegraphics[width=0.9\columnwidth]{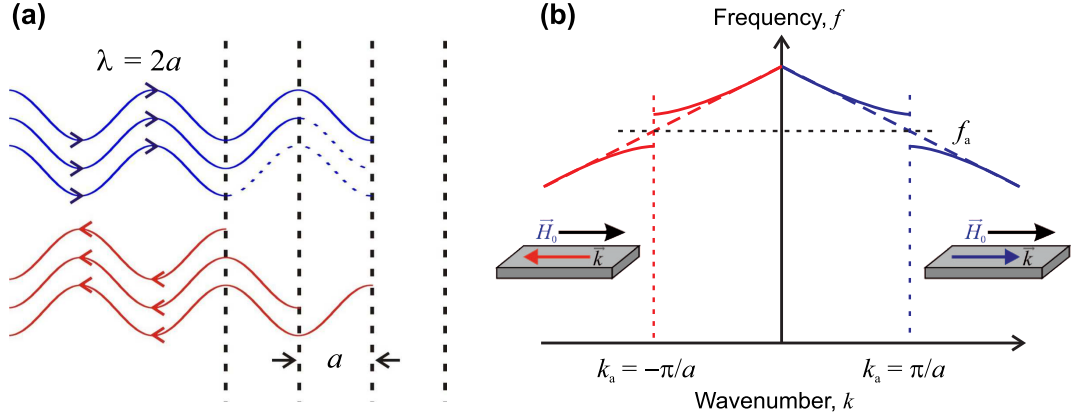}
\caption{\label{Fig3} (a) Sketch of Bragg reflection of a spin wave having an incident angle $\theta = \pi/2$ from a one-dimensional magnonic crystal. The Bragg law for the particular case shown in the figure is defined. (b) Schematic of the BVMSW modes dispersion curves in a uniform magnetic film (dashed lines) and in a magnonic crystal (solid lines). Formation of a band gap at the frequency corresponding to a spin-wave wavenumbers of $\pm \pi/a$ is shown.}
\end{center}
\end{figure}

Figure 3a shows schematically the mechanism of Bragg scattering in a one-dimensional magnonic crystal in the form of a periodic array of reflectors with a lattice constant $a$. The incident spin-wave angle $\theta$  in this case is equal to $\pi/2$. In most of the experiments, the reflection efficiency of the wave from a single reflector is rather small and varies in the range from 1 $\%$ to 10 $\%$ \cite{A9-chumak-2009-APL-muMC, A12-obry-2013-APL-ionMC, A6-chumak-2008-APL-MC} while the number of reflectors is usually limited to 20 due to a relatively large spin-wave loss \cite{A1-serga-2010-JPD-YIG}. In such a situation, the majority of the spin-wave energy propagates though the crystal with small parasitic loss if the Bragg condition $n \lambda = 2 a \cdot \sin \theta$  is not satisfied ($n$ is an integer value). As opposite, when the Bragg condition is satisfied, as it is shown in the figure, the spin waves that are reflected from each dashed line have the same phase providing a resonant reflection condition. The major part of the energy of such a spin wave will be reflected from the magnonic crystal. The dashed lines in Fig.~3b show simplified linearized dispersion curves for BVMSW modes propagating in an infinitely extended film along the biasing magnetic field in both directions. If the spin wave propagates through a magnonic crystal, the waves with wavenumber $k_a = \pm n \pi/a$ satisfying the Bragg condition are reflected resulting in the formation of a bang gap at the frequency $f_a$ (for $n = 1$) as it is schematically shown in the figure. The width of the band gap is usually defined by the reflection efficiency from a single reflector \cite{A20-karenowska-2012-PRL-DMC}.

As it was mentioned previously, spin-wave properties can be tuned by a variety of different parameters such as the thickness of the magnetic film, the width of the spin-wave waveguide, the saturation magnetization of the magnetic material or the applied biasing magnetic field \cite{2-gurevich, 3-stancil2}. This opens many possibilities for the realization of magnonic crystals, since each of these parameters can be varied periodically in space. If parameters are constant over time (e.g. they are defined only by the sample geometry), these magnonic crystals are termed static. Schematics of different types of static one-dimensional magnonic crystals together with their measured characteristics are shown in Fig.~4.

\begin{figure}
\begin{center}
\includegraphics[width=0.9\columnwidth]{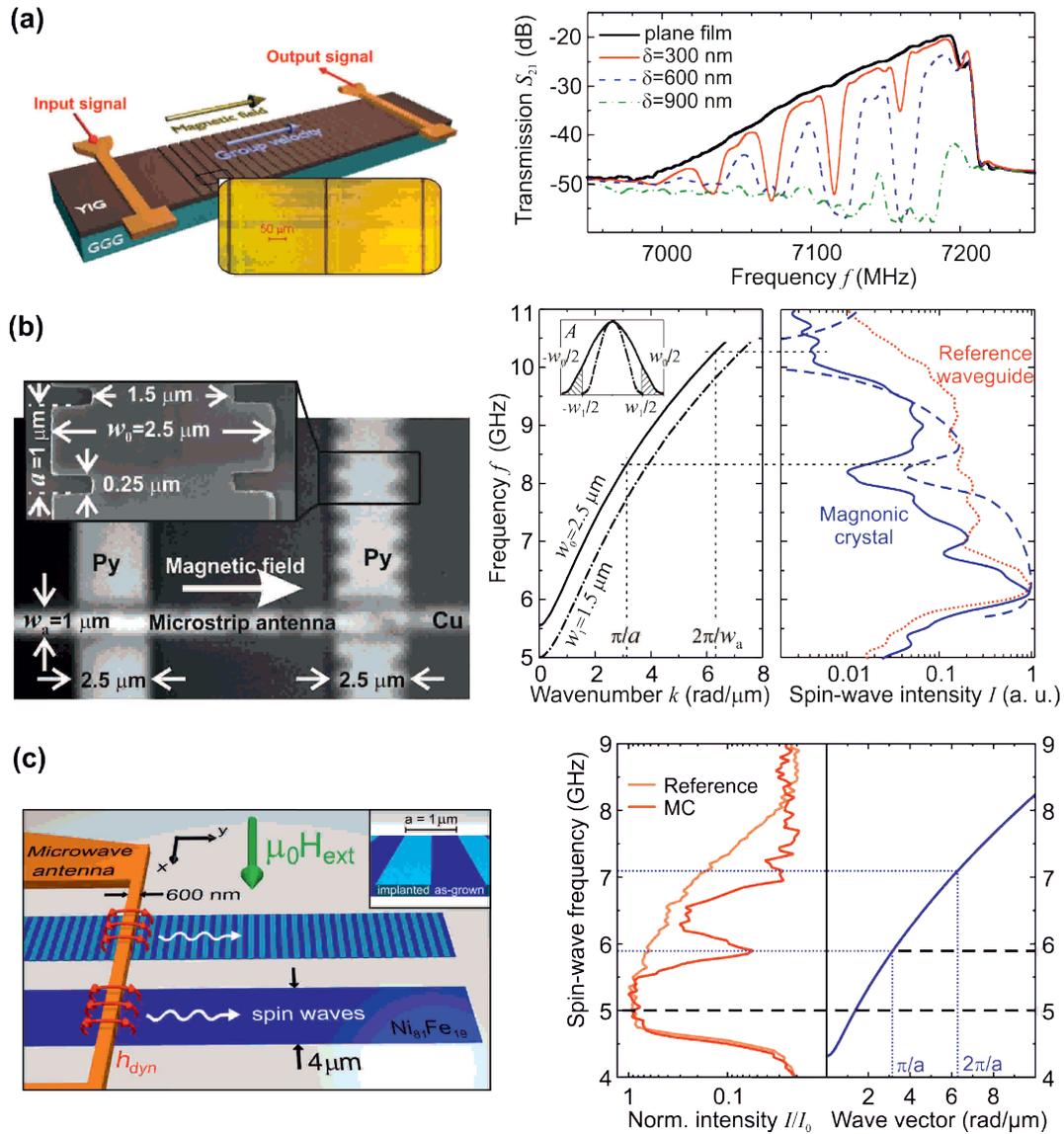}
\caption{\label{Fig4} Different realizations of static magnonic crystals. (a)  Sketch of a magnonic crystal structure comprising an array of shallow grooves on a surface of a YIG film. BVMSW transmission characteristics for magnonic crystals for different groove depths $\delta$ are shown \cite{A6-chumak-2008-APL-MC}. (b) Scanning electron microscopy and optical images of a width-modulated micro-scaled magnonic crystal together with a reference unpatterned waveguide. Calculated spin-wave dispersion curves and measured spin-wave intensity for magnonic crystal and reference waveguide are shown on the right \cite{A9-chumak-2009-APL-muMC}. (c) Schematic of the experimental setup for studies of micro-structured ion-implanted magnonic crystal \cite{A12-obry-2013-APL-ionMC}. Normalized spin-wave transmission spectra of the magnonic crystal waveguide (dark line) and the reference waveguide (pale line) as well as calculated spin-wave dispersion curves are shown on the right.}
\end{center}
\end{figure}

\subsection{Yttrium-iron-garnet magnonic crystals based on the periodic variation of film thickness}

Geometric structuring of a uniform spin-wave waveguide by fabricating an array of grooves on its surface is a simple and highly efficient means of magnonic crystal fabrication \cite{MC-review1, 17-MC-grooves1, A6-chumak-2008-APL-MC, A7-chumak-2009-APL-MC, A8-chumak-2009-JAP-MC} (see Fig.~4a). Photolithographic patterning followed by hot orthophosphoric acid etching (this etching technology utilizing the procedure of ultraviolet hardening of resist was developed at the Nano-Structuring Center of the University of Kaiserslautern \cite{A6-chumak-2008-APL-MC, A7-chumak-2009-APL-MC}) has been shown to be a reliable means of forming the required grooves in a YIG spin-wave waveguide. Spectra of transmitted spin waves in such structures have been studied experimentally and theoretically for backward volume \cite{A6-chumak-2008-APL-MC} (see right panel in Fig.~4a) and surface magnetostatic waves \cite{A7-chumak-2009-APL-MC}. Special attention has been paid to the study of the magnonic crystal microwave characteristics as a function of groove depth \cite{A6-chumak-2008-APL-MC, A7-chumak-2009-APL-MC, A8-chumak-2009-JAP-MC}. Pronounced rejection frequency bands have been clearly observed (more than 30 dB rejection was achieved) and it was shown that the rejection efficiency and the frequency width of the rejection bands increase with increasing groove depth (right panel in Fig.~4a). In addition, it has been found that the rejection of the BVMSW mode is considerably (up to 600 times) larger than of the MSSW modes. This was related to the non-reciprocal nature of the MSSW mode as well as to resonant scattering of the lowest BVMSW mode into higher thickness modes \cite{A7-chumak-2009-APL-MC}. 

Detailed studies of this magnonic crystal and its optimization have been performed in \cite{A8-chumak-2009-JAP-MC}. It was shown that the efficiency of the rejection can be controlled not only by the groove depth, but also by its width and by the number of grooves in the crystal. When the width of grooves is much smaller than the spin-wave wavelength, increasing the groove width leads to a fast increase in the rejection efficiency. The efficiency of the rejection also increases with an increase in the number of grooves. It was further found that the optimal groove depth which ensures strong rejection in the rejection bands while maintaining a low insertion loss in the transmission bands of around 3 dB is approximately 1/10 of the total film thickness for the BVMSW mode. Decreasing the groove depth further leads to an increase in the rejection efficiency in the band gap regions as well as in an increase in the parasitic loss in the transmission bands \cite{A8-chumak-2009-JAP-MC}. 

A theoretical model was developed to describe spin-wave transmission through a magnonic crystal \cite{A6-chumak-2008-APL-MC}. The model is based on introduction of transmission T-matrices for spin waves at spatial points where the thickness of the YIG waveguide is changing (steps), as well as for the sections of the waveguide with constant thicknesses \cite{A8-chumak-2009-JAP-MC}. As a result, the T-matrix for each periodic segment of a magnonic crystal was found as a multiplication of four different matrices two for the steps and two for the sections in between. In order to obtain the transmission characteristics for the whole magnonic crystal, this matrix was raised to the power equal to the number of periods. The results of this simple theoretical model have shown good qualitative agreement with the experimental results. Subsequently, this model was adopted to describe magnonic crystals with smooth transformations in magnetic properties \cite{A13-vogel-2015-NP-heatMC}.

\subsection{Micro-structured metallic magnonic crystals based on the periodic variation of the spin-wave waveguide width}

The groove-based magnonic crystals discussed above have macroscopic sizes (a typical lattice constant is 300 $\mu m$). Modern signal processing applications, however, demand magnonic crystals of sub-micron sizes. A nano-scaled magnonic crystal proposed in Ref.~\cite{width-1} has been fabricated from a Permalloy waveguide by engineering periodic variations in its width (see Fig.~4b) \cite{A9-chumak-2009-APL-muMC}. This system was studied experimentally using micro-focused Brillouin light scattering spectroscopy. A spin-wave band gap was clearly observed (see right panel in Fig.~4b). It is seen as a considerable decrease in the spin-wave transmission caused by the resonant backscattering from the periodic structure. The band gap frequency was tuned in the range from 6.5 to 9 GHz by varying the applied magnetic field. Results presented in \cite{A9-chumak-2009-APL-muMC} are the first experimental realization of the spin-wave propagation through a micro-structured magnonic crystal. The spin waves were exited here by a microstrip antenna deposited onto the top of the Permalloy waveguide. Another interesting method of magnon injection to the width-modulated magnonic crystals from a semi-infinite magnetic film was proposed and experimentally verified in Ref.~\cite{width-2}.

The transmission characteristics of Permalloy-based width-modulated crystals were also investigated by numerical simulations \cite{A10-ciubotaru-2012-APL-edgeMC, width-3}. The magnonic crystal was represented by a micro-sized planar ferromagnetic waveguide with periodically changing width. By choosing a step-like or sinusoidal variation of the width, the magnonic crystal revealed multiple or single band gaps, respectively \cite{A10-ciubotaru-2012-APL-edgeMC}. In addition, it was found that both the band gap frequency and its depth depend strongly on the probing position inside the magnonic crystal due the non-uniform distribution of the internal magnetic field. The dependence of the spin-wave spectrum on the position inside the magnonic crystal was also studied in Ref.~\cite{width-4}.

\subsection{Micro-structured metallic magnonic crystals based on the periodic variation of the saturation magnetization}

A micromagnetic analysis of spin-wave propagation was performed for a magnonic crystal realized as a Permalloy spin-wave waveguide with a periodical spatial variation of its saturation magnetization \cite{A11-ciubotaru-2013-PRB-ionMC}. Frequency band gaps were clearly observed in the spin-wave transmission spectra, and their origin is traced back to an overlap of individual band gaps of the fundamental and higher-order spin-wave width modes. This superposition could be controlled by the width of the magnonic crystal waveguide. Furthermore, the depths of the rejection bands depend strongly on both the level of the magnetization variation and the size of the area with a reduced magnetization: The reduction of the saturation magnetization over a larger area leads to the formation of more pronounced band gaps. 

As a proof of concept, a micro-structured Permalloy-based magnonic crystal was fabricated by localized ion implantation (see Fig.~4c) and was investigated in comparison with a reference Permalloy waveguide by means of Brillouin light scattering microscopy \cite{A12-obry-2013-APL-ionMC}. The irradiation caused a periodic variation in the saturation magnetization along the waveguide. It has been found that a relatively weak modification of the saturation magnetization by 7 $\%$ is sufficient to decrease the spin-wave transmission in the band gaps by a factor of 10 (see right panel in Fig.~4c). Since the waveguide consisted of one single, topographically unchanged material, spin waves with frequencies in the transmission bands were nearly unaffected. These results prove the applicability of localized ion implantation for the fabrication of efficient micron- and nano-sized magnonic crystals for magnon spintronic applications.

\section{Reconfigurable, dynamic, and moving magnonic crystals}

Reconfigurable magnonic crystals, whose properties can be changed on demand \cite{27-MC-review, 103-reconfMC, reconfig-1, reconfig-2, reconfig-3, reconfig-4, reconfig-5, reconfig-6, reconfig-7}, attract special attention since they allow for tuning of the functionality of a magnetic element: the same element can be used in applications as a magnon conduit, a logic gate, or a data buffering element. An example of such a structure is a magnonic crystal in the form of an array of magnetic strips magnetized parallel or anti-parallel to each other \cite{103-reconfMC}. The chosen magnetization state defines the lattice constant of the crystal and, thus, the spin-wave dispersion. 

From the point of view of small energy consumption, voltage-controlled reconfigurable magnonic crystals are of particular interest. A nano-scale reconfigurable magnonic crystal designed using voltage-controlled perpendicular magnetic anisotropy (PMA) in ferromagnetic-dielectric hetero-structures was investigated using numerical simulations in Ref. \cite{Wang-DMC}. A periodic array of gate metallic stripes have been placed on top of a MgO/Co structure in order to apply a periodic electric field and to modify the PMA in Co. It is demonstrated that the introduction of PMA modifies the spin-wave propagation and leads to the formation of band gaps in the spin-wave spectrum. The band gaps characteristics are defined by the applied electric field and can be controlled dynamically, i.e. it is possible to switch band gaps ON and OFF within a few tens of nanoseconds.

\subsection{Optically-induced reconfigurable magnonic crystals}

An important step towards the realization and studies of reconfigurable magnonic crystals was the demonstration that any two-dimensional magnetization pattern in a magnetic film can be created in a reconfigurable fashion by laser-induced heating \cite{A13-vogel-2015-NP-heatMC}. By using a laser, a thermal landscape in a magnetic medium was created (see Fig.~5a) that resulted in an equivalent landscape of the saturation magnetization \cite{104-obry-gradient} and, hence, in the control of the spatial spin-wave characteristics. The setup used for the realization of the light patterns consisted of a continuous-wave laser as light source, an acousto-optical modulator for temporal, and a spatial light modulator for spatial intensity control. In order to study the influence of the thermal gradient induced by the intensity patterns on the spin waves, we used a ferrimagnetic, 5 $\mu$m-thick YIG waveguide grown on a 500 $\mu$m thick GGG substrate (see Fig.~5a). GGG is almost transparent, while YIG absorbs about 40 $\%$ of the green light used in the experiment. In order to increase the efficiency of the heating we used a black absorber coated on top of the YIG film. The proposed concept of a reconfigurable magnetic material was demonstrated and tested on the example of one- and two-dimensional magnonic crystals. The formation of band gaps in the spin-wave transmission characteristic is shown in the right panel of Fig.~5a. It was demonstrated that the positions of the band gaps can be tuned by the lattice constant of the magnonic crystal, while the widths of the bandgaps were controlled by the laser light intensity \cite{A13-vogel-2015-NP-heatMC}.

\subsection{Current controlled dynamic magnonic crystals}

Changes in the properties of a magnonic crystal open access to novel physics if these changes occur on a timescale shorter than the spin-wave propagation time through the crystal. Such magnonic crystals are termed dynamic magnonics crystals \cite{ A15-chumak-book-2012-DMC, A14-chumak-JPD-DMC}. The first dynamic magnonic crystal was realized employing a YIG waveguide placed in a spatially periodic dynamically controllable magnetic field provided by a current carrying planar metallic meander structure (Fig.~5b) \cite{A14-chumak-JPD-DMC}. The current carrying structure was spaced at a distance of 100 $\mu$m above the YIG surface in order to avoid undesirable interaction between the spin waves and the conductor \cite{fetisov}. Since the lateral variations of the bias magnetic field were approximately sinusoidal inside of the YIG film, the spectrum of the crystal contained only one space component and a sole rejection band appears (see right panel in Fig.~5b). It was demonstrated that the rejection band depth and width could be tuned via the applied current \cite{A15-chumak-book-2012-DMC, A14-chumak-JPD-DMC}. Furthermore, it was shown that such a crystal could be switched from a full transmission to a full rejection state faster than 10 ns, which is ten times smaller than the spin-wave relaxation time and spin-wave propagation time through the crystal. The possibility of achieving fast temporal variation of crystal parameters opened doors to the study of new physical effects discussed in the following Section. 

\begin{figure}
\begin{center}
\includegraphics[width=0.9\columnwidth]{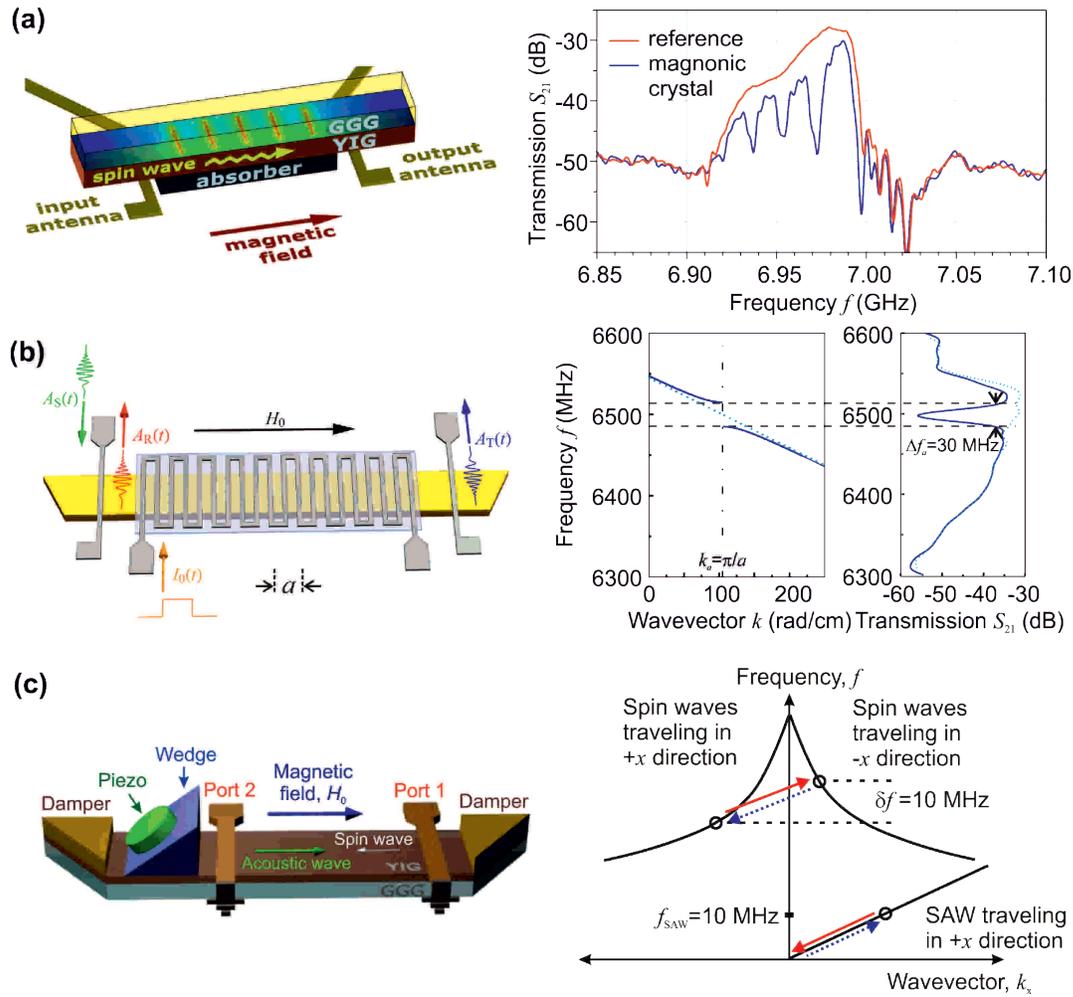}
\caption{\label{Fig5} Reconfigurable (a), dynamic (b), and travelling (c) magnonic crystals. (a) Schematic of the reconfigurable magnonic crystal consisting of a GGG/YIG/absorber multilayer system \cite{A13-vogel-2015-NP-heatMC}. A typical thermal landscape is shown on the waveguide (color coded: red - maximum temperature, blue - minimum temperature). Right panel: Measured spin-wave transmission characteristic (solid blue line) in the thermal landscape (magnonic crystal with five periods, lattice constant 740 $\mu$m) and reference data (red) without the projected pattern. (b) Schematic of the dynamic magnonic crystal comprising a planar current-carrying meander structure with 20 periods with a lattice constant a = 300\,$\mu$m (10 shown), positioned close to the surface of the YIG spin-wave waveguide. Right panel: Dispersion characteristic and measure a spin-wave transmission characteristic of this dynamic magnonic crystal for 1\,A current sent through the meander structure \cite{A14-chumak-JPD-DMC}. (c) Schematic of the moving magnonic crystal. Spin waves are excited and detected in the YIG film by stripline antennas. The surface acoustic wave is excited on the YIG/GGG sample by a piezoelectric quartz crystal and an acrylic wedge transducer \cite{A17-chumak-2010-PRB-SAW}. Right panel: Schematic of the dispersion curves for BVMSW and SAW. Circles indicate the waves that participate in Bragg scattering.}
\end{center}
\end{figure}

Another approach for the realization of a dynamic magnonic crystal was proposed in \cite{A16-nikitin-APL-2015-DMC} and is based on the control of the effective geometry of a magnetic film waveguide for spin waves via electric current. The device utilizes a spin-wave waveguide fabricated from a YIG waveguide of periodically-varied width and two conducting wires attached to the film surface along the modulated edges. A spatially uniform bias magnetic field is directed across the YIG waveguide. Positive currents applied to both wires induce an additional negative Oersted field that reduces the bias magnetic field near the edges. The magnetic wells which are thus created screen the YIG-film modulated edges from spin waves and, hence, suppress the magnonic crystal band gap. Advantages of the width-modulated dynamic magnonic crystal in comparison to the ones utilizing the meander wiring \cite{A14-chumak-JPD-DMC} are the potentially fast performance and the possibility for miniaturization. Indeed, the current control is provided here by short conductors, which have an inductance much less than the meander type wire. The use of nanostructured width-modulated waveguides made of Permalloy films (Fig.~4b), as in Ref. \cite{A9-chumak-2009-APL-muMC}, will allow for a significant reduction of the magnonic crystal size. This type of dynamic magnonic crystal has been used to perform AND logic gate operation \cite{A16-nikitin-APL-2015-DMC}.

\subsection{Surface acoustic wave based moving magnonic crystals}

Moving magnonic crystals represent a special class of dynamic and tunable crystals utilizing moving Bragg gratings. Such magnonic crystals have been created via a periodic strain induced by a surface acoustic wave (SAW) traveling along a YIG spin-wave waveguide as shown in Fig.~5c \cite{105-SAW-1, A17-chumak-2010-PRB-SAW, 106-SAW-2, kryshtal-2015}. Spin-wave scattering in such a crystal differs conceptually from the previously discussed cases since it takes place with a shift in the spin-wave frequency due to the Doppler effect. In fact, a combination of Bragg scattering with Doppler shift took place in the experiments \cite{A17-chumak-2010-PRB-SAW}. The schematic of such scattering is shown in the right panel of Fig.~5c. One can see, that in order to fulfill the conservation laws, the magnon scattering should occur with a shift in the frequency equal to the frequency of the surface acoustic wave. Simultaneously, the wavelengths of the spin waves and SAW should be comparable in order to satisfy momentum conservation.

The experiment was performed using backward volume magnetostatic waves which are characterized by a negative group velocity \cite{31-damon-eshbach2}. As a result, spin waves scattered from an approaching grating were found to be shifted down in frequency demonstrating the reverse Doppler effect \cite{A17-chumak-2010-PRB-SAW}. In contrast to Ref. \cite{107-doppler}, here the reflection occurs from a crystal lattice rather than from a single reflecting surface and, thus, the wave number of the scattered wave is strictly determined by the law of conservation of momentum. Due to this fact, the frequency-shifted wave appears as a single narrow peak in the transmission characteristic of the magnonic crystal. Apart from the interesting physics shown by such systems they possess potential for sensors and signal processing applications.

\section{Magnonic crystal based data processing}
\subsection{Magnonic crystal as a microwave filter}

One of the initial motivations in the study of spin-wave propagation through periodic magnetic structures was the realization of microwave filters and resonators for the processing of analogue information - see review \cite{MC-review1}. The magnonic crystals described above can also be used for these purposes. One can see in the spin-wave transmission characteristics shown in Figs.~4 and 5 that spin-wave spectra of a magnonic crystal comprise regions of frequencies for which the transmission of microwave signals is prohibited. Thus, a magnonic crystal can serve as a stop-band filter. The attenuation of the microwave signal in the stop band exceeded 30\,dB in our experiments \cite{A6-chumak-2008-APL-MC}, but was limited by the experimental conditions. Theoretical modelling has shown that this value might actually be much greater and can exceed 100\,dB. 

The advantage of such filters is that the central frequencies of the stop bands can easily be controlled by an applied magnetic field \cite{A9-chumak-2009-APL-muMC, A12-obry-2013-APL-ionMC} or by the lattice constant of the magnonic crystal \cite{A13-vogel-2015-NP-heatMC}. The width of the band gap, which is usually close to several tens of MHz is defined by the strength of the modulation of a magnetic property (e.g. groove depth \cite{A6-chumak-2008-APL-MC}, magnetic field \cite{A14-chumak-JPD-DMC}, the saturation magnetization \cite{A12-obry-2013-APL-ionMC}, mechanical stress \cite{A17-chumak-2010-PRB-SAW} or temperature \cite{A13-vogel-2015-NP-heatMC}) or by the thickness of the magnetic film that defines the slope of the spin-wave dispersion \cite{A8-chumak-2009-JAP-MC}. The number of band gaps in spin-wave spectra is given by the variation function of the magnetic properties \cite{A10-ciubotaru-2012-APL-edgeMC} (step-like functions result in multiple band gaps \cite{A6-chumak-2008-APL-MC}, harmonic variation of properties result in a single band gap \cite{A14-chumak-JPD-DMC}). Moreover, one of the greatest advantages of such filters is determined by the fact that the spin-wave wavelengths are orders of magnitude smaller compared to the wavelengths of electromagnetic waves in the same GHz frequency range. That allows for a considerable miniaturization of the devices (e.g. magnonic crystals of micro-meter sizes were realized \cite{A9-chumak-2009-APL-muMC, A12-obry-2013-APL-ionMC}). Finally, reconfigurable and dynamic magnonic crystals allow for the realization of microwave filters that can be tuned or switched ON/OFF on a nanosecond time scale \cite{A14-chumak-JPD-DMC}.

On the other hand, the main drawbacks of microwave filters based on magnonic crystals are probably the relatively low efficiencies of spin-wave excitation and detection by standard micro-strip antennas and relatively high values of spin-wave damping. This induces loss in the transmission bands of the filter which, in our experimental demonstrators, usually exceed 10\,dB (see Figs.~4 and 5). By utilizing thicker magnetic films and optimized microwave antennas this value can be improved but will probably still be close to a few dB \cite{MC-review1, MC-review2}. 

\subsection{Microwave signal generation based on magnonic crystals with a feedback 
loop}

Spin-wave systems as discussed above also allow the realization of coherent microwave sources. Self-exciting positive-feedback spin-wave systems, often termed spin-wave active rings, are used for this purpose \cite{108-loop-1, 109-loop-2}. The basis for such an active ring is a dispersive spin-wave waveguide with exciting and receiving antennas connected together via a variable-gain electrical feedback loop (see Fig.~6a). If the correct gain and phase conditions are met, a monochromatic (but noise-initiated) signal propagates in the ring, increases with time until nonlinear saturation takes place either in the spin-wave system or in the external amplifier. The signal from the active ring can be partially extracted using a directional coupler as shown in the schematic.

\begin{figure}
\begin{center}
\includegraphics[width=0.9\columnwidth]{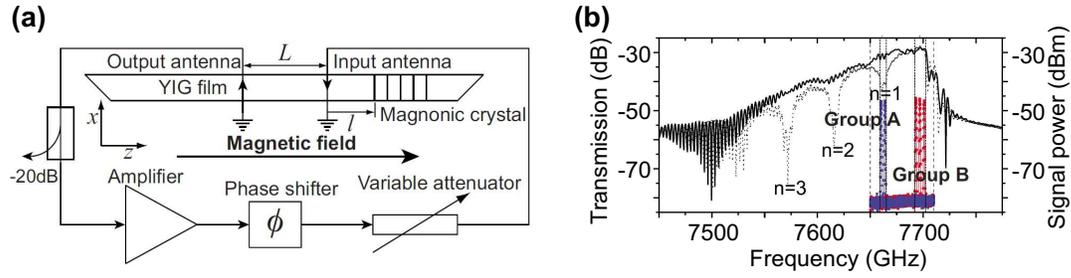}
\caption{\label{Fig6} (a) Schematic of an experimental active ring system employing a magnonic crystal to define the wavenumber of the resonance mode. Two microstrip antennas excite and receive spin-waves through a region of the YIG film. A magnonic crystal, which is part of the same film, is located in close proximity to the input antenna. (b) BVMSW transmission characteristics for the uniform (bold, solid) and magnonic crystal (dotted) regions of the YIG film (left ordinate axis). Inset diagram (right ordinate axis) is a composite of five overlaid ring power spectra for different distances between the input antenna and the magnonic crystal \cite{A18-karenowska-2010-APL-Gener}. }
\end{center}
\end{figure}

Spontaneous excitation of an active ring system occurs if the external gain is increased beyond the self-generation threshold of the spin-wave mode with the smallest threshold. This mode is called dominant and suppresses the excitation of all others modes in the quasi-linear regime. However, the wavenumber of this dominant mode, which defines the generated frequency, is highly sensitive to the internal magnetic field and the thermal environment of the spin-wave transmission medium. As a result, in any real system it is almost impossible to predict the wavenumber of the dominant mode. It was demonstrated however that exploitation of spin-wave reflections from a groove-based magnonic crystal allows for the selection of the wavenumber of the dominant mode \cite{A18-karenowska-2010-APL-Gener}. 

The active ring shown in Fig.~6a operates with BVMSWs that are reciprocal. This means that spin waves excited by the input antenna propagate in both directions of the waveguide. In order to achieve an auto-oscillation regime of the ring, constructive interference between the signal detected at the output antenna, amplified and fed back to the input antenna, and the signal reflected from the magnonic crystal (see schematic) is required. The tuning of the phases of both signals plays a crucial role in this case. The phase of the signal from the output antenna is tuned by the external phase shifter, while the phase of the signal reflected from the magnonic crystal is defined by the distance between the input antenna and the first groove of the crystal. Thus, by changing the position of the crystal relative to the antennas, different dominant modes were excited in the experiment (see Fig.~6b). Excitation of two different groups of modes was observed. Group B corresponds to the conventional operating regime of the active ring in which the magnonic crystal does not play any role. It occurs at frequencies where the spin-wave transmission is close to its maximum. In contrast, the wavenumbers of the spin waves defining group A are determined by the magnonic crystal characteristics and lie within the first band gap. The position of the magnonic crystal relative to the antennas defined which of the three modes inside the band gap were generated. Thus, when mode enhancement phase conditions were satisfied, the ring geometry permits highly robust forced dominant wavenumber selection \cite{A18-karenowska-2010-APL-Gener}. 

Furthermore, it was demonstrated in Ref.~\cite{loop-2} that a delay line based on a one-dimensional magnonic crystal used in a feedback loop of a microwave auto-oscillator can substantially reduce the phase noise figure and improve other vital performance characteristics of the auto-oscillator. 
It is also interesting that the frequency selectivity of a magnonic crystal with quasiperiodic Fibonacci type structure may provide conditions for the self-generation of dissipative solitons in an active spin-wave ring \cite{loop-1, loop-3}.

\subsection{Frequency inversion and time reversal using dynamic magnonic crystals}

In the previous Section, the transmission properties of our proof-of-concept dynamic magnonic crystal was discussed when it is kept continuously either in the OFF or the ON state (see Fig.~5b). In follow-up studies, it was demonstrated that dynamic magnonic crystals open access to new physical effects if the crystal is switched from OFF to ON whilst a spin-wave packet is inside \cite{A20-karenowska-2012-PRL-DMC, A19-chumak-2010-NC-TimeRev}. 

If a spin wave with wavevector $k_s \approx +k_a = +\pi/a$ is incident on the dynamic magnonic crystal whilst it is in the ON state, it is reflected with conservation of the frequency. However, if the crystal is switched from OFF to ON whilst such a wave is inside, the situation is quite different \cite{A19-chumak-2010-NC-TimeRev}. In Fig.~7a dispersion curves for incident signal waves (green section) and those reflected by the magnonic crystal (red section) are shown. Black dots mark the reference frequency $f_a$ lying in the center of the bandgap and corresponding to the Bragg wave vectors $\pm k_a = \pm \pi/a$. The green open circle and square illustrate two spectral components of the incident signal waveform. The spatially periodic magnetic modulation of the waveguide's magnetic bias field, brought about by the application of the current pulse to the dynamic magnonic crystal meander structure (see Fig.~5b), couples these components to corresponding components of the reflected waveform (red open circle and square). The difference between the wave vectors of the signal and reflected waves is fixed by the lattice constant $a$ of the dynamic magnonic crystal such that the $k$-spectrum of the reflected waveform is uniformly shifted to the left by $\Delta k = 2 \pi/a$ (lower panel). This shift results in a spectral inversion in the frequency domain. The frequency inversion in a range of frequencies of $\pm$15 MHz has been shown experimentally \cite{A15-chumak-book-2012-DMC, A20-karenowska-2012-PRL-DMC, A19-chumak-2010-NC-TimeRev}.

\begin{figure}
\begin{center}
\includegraphics[width=0.9\columnwidth]{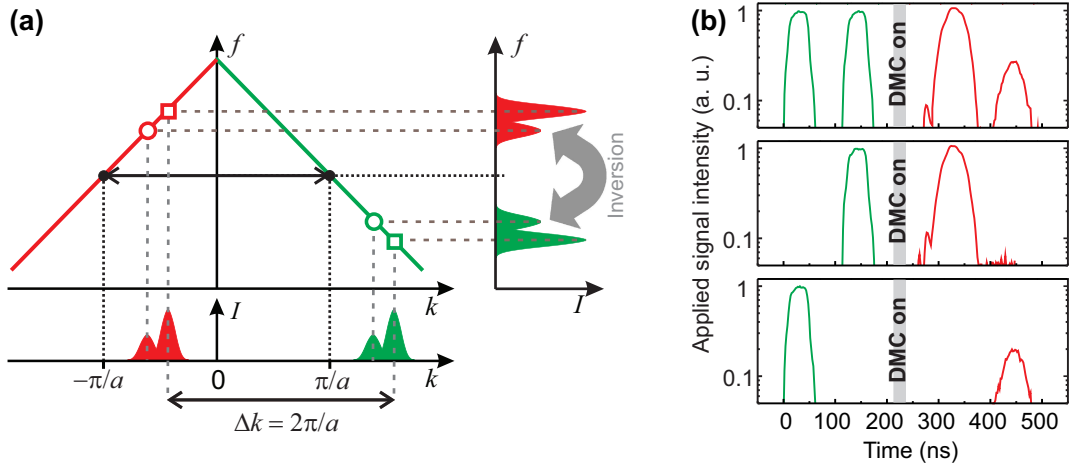}
\caption{\label{Fig7} Frequency inversion and time reversal by a dynamic magnonic crystal \cite{A19-chumak-2010-NC-TimeRev}. (a) Schematic of the frequency inversion process. (b) The experimentally detected spin-wave signals reflected by the dynamic magnonic crystal. The gray area shows the time during which the dynamic magnonic crystal was switched ON. }
\end{center}
\end{figure}

 Time-reversing a signal is equivalent to inverting its spectrum about a reference frequency. This can be easily proven by considering the Fourier domain description of an arbitrary signal envelope which shows that the frequency inversion is equivalent to the transformation $t \rightarrow -t$. In order to demonstrate the spin-wave time reversal a BVMSW packet consisting of two pulses was used -- see Fig.~7b. Switching OFF of the first or the second pulse has clearly demonstrated that the spin-wave packet reflected by the magnonic crystal is reversed in time: In the case that two spin-wave pulses are applied, two corresponding pulses reflected by the dynamic magnonic crystal are observed (upper frame, red). When the first of the two pulses is switched OFF, the second reflected pulse is absent (middle frame), and vice versa (lower frame), confirming time reversal \cite{A19-chumak-2010-NC-TimeRev}. 

Thus, it was shown that a dynamic magnonic crystal may provide a linear means to perform spectral transformations, including frequency inversion and time reversal, which, until now, have only been possible through nonlinear mechanisms \cite{9-WFR-melkov2, 10-WFR_2D, 110-WFR-1, 111-WFR-2}. These results go far beyond spin waves since they can be applied to waves of any nature. For example, the idea of all-linear time reversal was taken up by Y. Sivan and J.B. Pendry working in the field of photonics \cite{112-sivan}.

\subsection{Data buffering in standing magnonic crystal modes}

The deceleration or even the full stop of light due to the modification of the light dispersion in photonic crystals has been a topic of intense studies over the last decade \cite{113-krauss, 114-baba}. A wave of light propagating through a photonic crystal couples with the internal standing crystal mode and generates a slow light mode which can be used for storage of optical signals. Magnonic crystals as a magnetic counterpart of photonic crystals operate with spin waves and can demonstrate similar effects in the GHz frequency range. However, in spite of the progress in magnonic crystals studies, no significant spin-wave deceleration has been demonstrated. This is due to the pronounced spin-wave damping, which limits the maximum number of structure periods to 20 or so. The small number of periods implies that the group velocity of spin waves, rather than vanishing, only slightly decreases at the gap edges \cite{A21-chumak-2012-PRL-storage}. Experimental investigations of groove-based magnonic crystals have shown that the spin-wave velocity can be decreased by about 20 percent at the edges of band gaps \cite{A21-chumak-2012-PRL-storage}. 

\begin{figure}
\begin{center}
\includegraphics[width=0.9\columnwidth]{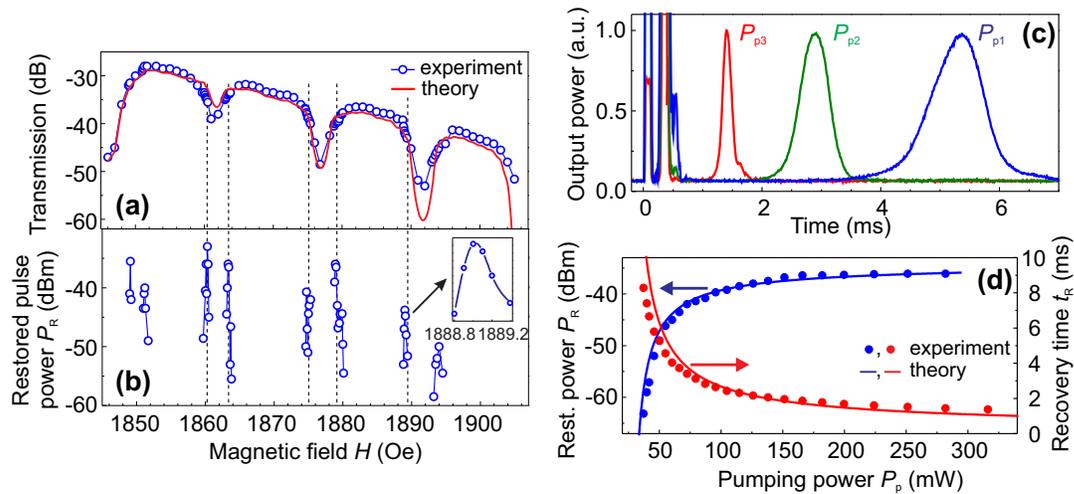}
\caption{\label{Fig8} Microwave signal restoration in a parametrically driven magnonic crystal \cite{A21-chumak-2012-PRL-storage}. (a) Transmission of the spin-wave signal as a function of bias magnetic field. (b) Power of the restored pulse as a function of field. One sees that the restored signal is visible only at the edges of the band gaps. (c) Time profiles of the restored signal measured at 1860 Oe magnetic field for different pumping powers: $P_{p1} =$ 50 mW, $P_{p2} =$ 90 mW, $P_{p3} =$ 320 mW. (d) Measured (circles) and calculated (lines) dependencies of the restored signal power and recovery time as a function of the pumping power.}
\end{center}
\end{figure}

Nevertheless, it was demonstrated that the storage of spin information in spin-wave modes is possible also in magnonic crystals. Unlike the conventional scheme used in photonics, this trapping occurs not due to the deceleration of the incident wave when it enters the periodic structure \cite{113-krauss, 114-baba} but due to the excitation of quasi-normal modes of an artificial crystal \cite{A21-chumak-2012-PRL-storage}. The first quasi-normal mode, whose eigen frequency coincides with the minima of the spin-wave group velocity at the edges of the magnonic crystals band gaps, has the longest lifetime among all the other modes and, thus, conserves the signal energy and phase information for a long time after the propagating wave has left the magnonic crystal. Such a trapped spin wave signal can subsequently be restored to its traveling form by means of double-frequency parametric amplification. One can see in Figs.\,8c and 8b that the restoration and, thus, the storage occurs only at the edges of the magnonic crystal band gap \cite{A21-chumak-2012-PRL-storage}.

Figure 8c shows time profiles of the restored signals for different pumping powers. The amplitude, the time of appearance and the duration of the restored signal were determined by the pumping power. The restoration mechanism based on a competitive nonlinear amplification process has been proposed earlier \cite{115-restor} and a corresponding theoretical model was developed \cite{A22-chumak-2009-PRB-storage} for the case of featureless thin magnetic films. This model describes the dependencies of the restored pulse amplitude, duration, and delay as a function of pumping power with high accuracy (see solid lines in Fig.~8d). The longest delay between the excitation of the data-carrying spin-wave pulse, which propagated through the magnonic crystal, and application of the pumping pulse to read out this information was 1.6 $\mu$m in the experiments \cite{A21-chumak-2012-PRL-storage}. Thus, the proposed physical phenomena can be used for short-time data buffering over a few microseconds but not for long time storage. 

In addition, investigations of parametric interactions between spin waves and electromagnetic pumping in magnonic crystals have delivered extremely interesting insight into nonlinear wave physics. In such systems, parametrically coupled waves propagating in opposite directions are simultaneously coupled by the Bragg reflection law. As a result, the parametric instability in magnonic crystals exhibits a strongly phase dependent behavior \cite{A21-chumak-2012-PRL-storage}.

\subsection{Magnonic crystals in spin-wave logic devices}

One of the strengths of magnonics lies in the benefits provided by the wave nature of magnons for processing of digital data and computation. Nowadays, new technologies, allowing for example the fabrication of nanometer-sized structures or operation in the THz frequency range, in combination with novel physical phenomena, provide new momentum to the field and make magnonic crystals to be especially promising for controlling and manipulating magnon currents \cite{A3-chumak-2015-NP-MagSpin}. The idea of encoding binary data into the spin-wave amplitude was first proposed in a theoretical study in Ref.\,\cite{116-logic0} and first experimental steps in this direction were made in Ref.\,\cite{117-logic1}. It was proposed to use a Mach-Zehnder spin-wave interferometer equipped with current-controlled phase shifters embedded in the interferometer arms to construct logic gates. Following this idea, the proof-of-principle XNOR and NAND logic gates were realized shortly thereafter \cite{118-logic2}. 

The dynamic magnonic crystal based on a width-modulated waveguide \cite{A16-nikitin-APL-2015-DMC} discussed above also can be used as a logic gate. The operating principle is based on controlling the bias magnetic field distribution along the sinusoidal borders of the YIG-film waveguide through a change in the electric currents $I_1$ and $I_2$. The logic "0" is represented by zero input current and a logic "1" by  the current which is enough to "switch OFF" the width modulation on one side of the magnonic crystal. The microwave pulses at the input antenna represent clock pulses. The microwave signal at the output antenna serves as the logic signal output. The operational frequency of the device corresponds to the central frequency of the band gap. Therefore, the presence of the band gap (i.e., low power level at the output) represents a logic "0" and the absence of the band gap represents a logic "1". It is proven experimentally that the logic "0" appears at the output port in three situations: for two logic "0s" applied to the input ports and for the combinations of the logic "0" and the logic "1" applied to the inputs. In the case where a logic "1" is applied simultaneously to both inputs, the band gap disappears, the spin-wave passes through the magnonic crystal, and the signal at the output port corresponds to a logic "1". Thus, the performance of the this dynamic magnonic crystal as an AND logic gate was demonstrated \cite{A16-nikitin-APL-2015-DMC}.

Alternatively, the spin-wave phase was proposed to be used for digitizing information instead of the amplitude (see review \cite{120-khitun1}). A wave with some chosen phase $\phi_0$ corresponds to a logic ``0'' while a logic ``1'' is represented by the wave with phase $\phi_0+\pi$. Such an approach allows trivial embedding of a NOT logic element in magnonic circuits by changing the position of the read-out device by a distance equal to half the wavelength. Moreover, it opens access to the realization of a majority logic gate in the form of a multi-input spin-wave combiner \cite{122-khitun3, 92-dutta, 121-khitun2,  A24-klingler-2014-APL-majGate, A25-klingler-2015-APL-MajGate, gonnenwein, braecher, MG-experiment}. The spin-wave majority gate consists of three input waveguides where spin waves are excited, a spin-wave combiner which merges the different input waveguides, and an output waveguide where a spin wave propagates with the same phase as the majority of the input waves. Furthermore, this device can perform not only majority operations but also AND and OR operations, if one of its inputs is used as a control input \cite{120-khitun1, A24-klingler-2014-APL-majGate}. Recently the experimental prototype of the three-input majority gate was realized and characterised \cite{MG-experiment}. It was proven that, in accordance with predictions from numerical simulations, the phase of the output signal represents the majority of the phase of the input signals. A switching time of about 10\,ns in the prototype evidences the ability of sub-nanosecond data processing in future micro-scaled devices.

The first micro-scale design of the majority gate was demonstrated using micromagnetic simulations in Ref.\,\cite{A24-klingler-2014-APL-majGate}. One of the main problems of a realistic spin-wave majority gate is the coexistence of different spin-wave modes with different wavelengths at a fixed frequency. By choosing the proper width of the output waveguide, it was possible to select the first width mode from the combiner and to ensure readability of the output signal. Still, the output signal in \cite{A24-klingler-2014-APL-majGate} was influenced by exchange spin waves of the same frequency but with much shorter wavelengths. To overcome these limitations, the isotropic forward volume magnetostatic spin waves (FVMSWs) were used in \cite{A25-klingler-2015-APL-MajGate}. Moreover, a high spin-wave transmission through the gate of up to 64$\%$, which is about three times larger than for the in-plane magnetized gate was achieved \cite{A25-klingler-2015-APL-MajGate}. A new asymmetric design of the majority gate has been proposed for this purpose. 
Another promising solution, which is under investigation at the moment, is the application of the magnon crystals for spin-wave mode selection in both in-plane and out-of-plane geometries.

An important advantage of the majority gate is that it might operate with spin waves of different wavelengths simultaneously, paving the way toward single chip parallel computing. This approach requires splitting of spin-wave signals having different wavelengths and, therefore, the usage of magnonic crystals is of crucial importance here \cite{122-khitun3}. 

\subsection{Magnon transistor}

The drawback of electric current-controlled spin-wave logic devices \cite{A16-nikitin-APL-2015-DMC,116-logic0,117-logic1,118-logic2,119-logic3} is that it is impossible to combine two logic gates without additional magnon-to-voltage and voltage-to-magnon converters. This fact stimulated the search for means to control a magnon current by another magnon current. Recently, it has been demonstrated that such control is possible owing to nonlinear magnon-magnon scattering, and a magnon transistor allowing all-magnon data processing has been realized (see Fig.~9) \cite{A23-chumak-2014-NC-transist}. In this three-terminal device, the density of the magnon current flowing from the source to the drain (see blue spheres in the inset of Fig.~9b) is determined by the amount of magnons injected into the gate of the transistor (red spheres). The magnonic crystal in the form of an array of surface grooves is used to increase the density of the gate magnons and, consequently, to enhance the efficiency of the nonlinear four-magnon scattering process used to suppress the source-to-drain magnon current. It was shown that the source-to-drain magnon current can be decreased by up to three orders of magnitude (see bottom panel in Fig.~9b) \cite{A23-chumak-2014-NC-transist}. Although the operational characteristics of the presented insulator-based transistor in its proof-of-principle-type form do not overcome those of semiconductor devices, the presented transistor might play an important role for future magnonics technologies, in which information will be carried and processed by magnons \cite{A3-chumak-2015-NP-MagSpin}.

For example, two magnon transistors embedded into the arms of the Mach-Zehnder interferometer allow to realize a XOR gate shown in Fig.~9c. Initially, the feeding magnon current, which is sent to the interferometer, is divided into two identical currents. These currents are injected to the transistors' sources $\mathrm{S_1}$ and $\mathrm{S_2}$ and are independently controlled by the input magnon signals $I_1=n_{\mathrm{G}_1}$ and $I_2=n_{\mathrm{G}_2}$ applied to the gates $\mathrm{G_1}$ and $\mathrm{G_2}$. The input signal $I=$``1'' corresponds to the critical gate magnon density $n_\mathrm{G}^\mathrm{crit}$ large enough to decrease the magnon density at the drain to the value $n_\mathrm{D}/n_\mathrm{D_0}\leq 1/3$ (see Fig.~9b); the input $I=$``0'' means the absence of the gate magnons (the transistor is open). 
In the case when both input signals are zero $I_1=I_2=$``0'', the output signal $O$ after the combiner has zero value ``0'' due to the destructive interference (see permanent $\pi/2$ phase shift which appears twice in Fig.~9c). 
The application of a signal to only one of the transistors switches OFF one of the magnon currents and therefore switches OFF the destructive interference resulting in $O=$``1''. Finally, switching OFF both currents results in the absence of magnons at the devices output $O=$``0'' (see the truth table in Fig.~9c). 
\begin{figure}
	\begin{center}
		\includegraphics[width=0.9\columnwidth]{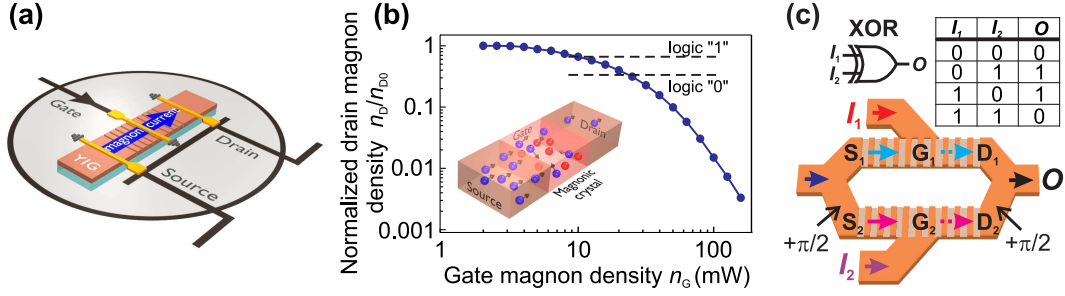}
		\caption{\label{Fig9} (a) Schematic of the magnon transistor \cite{A23-chumak-2014-NC-transist}. The transistor is based on a magnonic crystal designed in the form of a YIG film with an array of parallel grooves at its surface. The magnons are injected into the transistor's source and are detected at its drain using microstrip antennas. (b) The measured drain magnon density is presented as a function of the gate magnon density. The horizontal dashed lines define the drain density levels corresponding to a logic ``1'' and a logic ``0''. Schematic of the operational principle of a magnon transistor is shown in the inset. The source-to-drain magnon current (shown with blue spheres) is nonlinearly scattered by gate magnons (red spheres) injected into the gate region. 
		(c) Magnon crystal-based all-magnon chip proposed for XOR logic operation and the corresponding truth table.}
	\end{center}
\end{figure}

\section{Summary}

Magnonic crystals are artificial magnetic media with periodic variations of their magnetic properties in space. Bragg scattering affects the spin-wave spectrum in such a periodic structure and leads to the formation of band gaps---frequencies at which spin-wave propagation is prohibited. Consequently, areas between the bandgaps allow for selective spin-wave propagation, whereas the pronounced changes of the spin-wave dispersion near the bandgap edges cause the formation of bandgap solitons, the deceleration of spin waves, and the appearance of confined spin-wave modes. Magnonic crystals are rapidly gaining recognition as structures, which have a lot to contribute not only to the study and technological application of spin waves, but also to our general understanding of complex wave dynamics.
 
In the following, the main fundamental and applied scientific achievements presented in this article are summarized.

\begin{itemize}
  \item{A new technology based on photolithographic patterning followed by hot orthophosphoric acid etching was developed for the fabrication of groove-based YIG magnonic crystals \cite{A6-chumak-2008-APL-MC}. Propagation of backward volume magnetostatic waves in such crystals was studied for the first time and a high efficiency in the formation of band gaps was observed \cite{A6-chumak-2008-APL-MC}. The rejection efficiency and the width of the band gaps was studied systematically as a function of groove depth and width as well as the number of grooves.}
  \item{It was observed that the rejection efficiency of surface magnetostatic spin waves in groove based magnonic crystals is strongly suppressed (up to 600 times) compared to the scattering of backward volume magnetostatic waves \cite{A7-chumak-2009-APL-MC}. It was suggested that this is related to the non-reciprocal nature of the MSSW mode as well as to resonant scattering of the lowest BVMSW mode into higher thickness modes.}
  \item{A technology for the fabrication of YIG structures of micrometer sizes was developed and the spin-wave propagation in a micro-scaled YIG waveguide was investigated for the first time \cite{A4-pirro-2014-APL-microYIG}. The waveguide was fabricated using e-beam lithography with subsequent Ar$^{+}$ ion etching. The spin waves were detected by Brillouin light scattering spectroscopy. It was shown that the mean free path of spin waves in micro-structured YIG reaches at least 31\,$\mu$m, which is much longer than in analogous metallic structures.} 
  \item{A width-modulated micro-scaled Permalloy magnonic crystal was developed and studied \cite{A9-chumak-2009-APL-muMC}. The first experimental studies of the propagation of artificially excited spin waves through a micro-scaled magnonic crystal was performed. The formation of pronounced band gaps in the spin-wave spectrum was demonstrated.} 
  \item{It was shown, by means of numerical simulations, that the single or multiple band gap nature is defined by choosing a sinusoidal or step-like variation of the width of the spin-wave waveguide, respectively \cite{A10-ciubotaru-2012-APL-edgeMC}. This opens an additional degree of freedom in the control of the magnonic crystal properties.}
  \item{Ion implantation of a Permalloy spin-wave waveguide for the realization of a magnonic crystal with a periodic variation of the saturation magnetization was accomplished for the first time \cite{A12-obry-2013-APL-ionMC}. The appearance of clearly defined band gaps was shown which is due to the absence of parasitic excitation of higher-order width spin-wave modes in the waveguide. The spin-wave dynamics in such crystals was studied in detail using numerical simulations \cite{A11-ciubotaru-2013-PRB-ionMC}.}
  \item{Reconfigurable magnonic crystals based on two-dimensional laser-induced heating with corresponding magnetization patterns were realized \cite{A13-vogel-2015-NP-heatMC}. This approach, in contrast to all other existing techniques, opens access to full dynamic control of the magnetic properties in two dimensions.}
  \item{The first dynamic magnonic crystal was realized using an electric current-carrying meander structure placed in a vicinity of the spin-wave waveguide \cite{A14-chumak-JPD-DMC}. It was shown that the characteristics of the crystal are defined by the magnitude of the DC current. It was demonstrated that the crystal can be switched ON and OFF faster than 10\,ns which is up to 30 times faster than the spin-wave relaxation time and the spin-wave propagation time through the crystal.}
  \item{Another approach for the realization of a dynamic magnonic crystal based on a width-modulated spin-wave waveguide was proposed and used for the realization of a AND logic gate for processing of digital data \cite{A16-nikitin-APL-2015-DMC}.}  
  \item{Surface acoustic waves were used for scattering of spin waves and to realize a travelling magnonic crystal \cite{A17-chumak-2010-PRB-SAW}. Scattering of backward volume magnetostatic waves having negative group velocity was investigated and the reverse Doppler shift of the frequencies of the scattered waves was observed \cite{A17-chumak-2010-PRB-SAW}.}
  \item{A cumulative achievement is that it was demonstrated that practically any physical parameter that determines spin-wave properties can be varied periodically in space in order to realize a magnonic crystal. Particular parameters which were varied are: (i) spin-wave waveguide thickness \cite{A6-chumak-2008-APL-MC}, (ii) spin-wave waveguide width \cite{A9-chumak-2009-APL-muMC}, (iii) saturation magnetization of the waveguide \cite{A12-obry-2013-APL-ionMC}, (iv) mechanical stress \cite{A17-chumak-2010-PRB-SAW}, and (v) biasing magnetic field \cite{A14-chumak-JPD-DMC}.}
  \item{By utilizing the various accomplishments of our research it was shown that magnonic crystals can be used as a micro-scaled microwave filter with tunable characteristics \cite{A6-chumak-2008-APL-MC, A7-chumak-2009-APL-MC, A8-chumak-2009-JAP-MC, A9-chumak-2009-APL-muMC, A10-ciubotaru-2012-APL-edgeMC, A11-ciubotaru-2013-PRB-ionMC, A12-obry-2013-APL-ionMC, A13-vogel-2015-NP-heatMC, A14-chumak-JPD-DMC, A15-chumak-book-2012-DMC, A16-nikitin-APL-2015-DMC}.} 
  \item{A magnonic crystal was used as a resonator in order to set the wavenumber and thus to enhance the stability of the microwave signal generation based on a feedback loop system \cite{A18-karenowska-2010-APL-Gener}.} 
  \item{The coupling of two-counter-propagating spin-wave modes \cite{A20-karenowska-2012-PRL-DMC} as well as frequency inversion and time reversal were realized for the first time using a dynamic magnonic crystal \cite{A19-chumak-2010-NC-TimeRev}. These results are of importance for wave-based physics as a whole, since they demonstrate that typically non-linear wave phenomena can be realized in all-linear but dynamic systems.} 
  \item{The phenomenon of coherent wave trapping and restoration was demonstrated in a groove-based magnonic crystal \cite{A21-chumak-2012-PRL-storage}. Unlike the conventional scheme used in photonics, the trapping occurs not due to the deceleration of the incident wave when it enters the periodic structure but due to the excitation of quasi-normal modes of the artificial crystal. The restoration of the traveling wave was implemented by means of phase-sensitive parametric amplification of the stored mode.}
  \item{It is shown that width-modulated dynamic magnonic crystal can be used as a AND logic gate.}
  \item{Two different designs of a micro-scaled majority gate for processing of digital information were proposed and studied using numerical simulations \cite{A24-klingler-2014-APL-majGate, A25-klingler-2015-APL-MajGate} and experimentally \cite{MG-experiment}. Their functionality was proven. Magnonic crystals allow for the spin-wave filtering in such devices and therefore open access to parallel computing in which devices operate simultaneously with linear superposition of spin-waves of different wavelengths.} 
  \item{A groove-based magnonic crystal was used to enhance nonlinear magnon-magnon interactions, and for the realization of magnon-by-magnon control. This resulted in the development of the first magnon transistor \cite{A23-chumak-2014-NC-transist}. It was shown that the density of magnons flowing from the transistor's source to its drain can be decreased by three orders of magnitude through the injection of magnons into the transistor's gate connected to the magnonic crystal.}
 
\end{itemize} 

\section*{Acknowledgements}
The majority of our results presented in this article were obtained in the frame of the DFG project ``Linear and non-linear magnonic crystals'' (SE 1771/1) in close collaboration with B. L\"{a}gel and S. Wolff (Nano Structuring Center, University of Kaiserslautern), M. Vogel and G. von Freymann (Group of Optical Technologies and Photonics, University of Kaiserslautern), M. P. Kostylev (University of Western Australia), V. S. Tiberkevich and A. N. Slavin (Oakland University), A. D. Karenowska and J. F. Gregg (University of Oxford), P. Dhagat and A. Jander (Oregon State University), the groups of B. A. Kalinikos (St. Petersburg Electrotechnical University), and J. Fassbender (Helmholtz-Zentrum Dresden-Rossendorf). A. V. Chumak also acknowledges the financial support by ERC Starting Grant 678309 MagnonCircuits, A. A. Serga acknowledges the financial support by DAAD (Projekt 57213643 ``Linear and nonlinear spin-wave dynamics in magnonic-crystal-based micro-sized magnon conduits''), and B. Hillebrands acknowledges the financial support by the DFG in the frame of SFB/TRR 173: SPIN+X.

\end{document}